\documentclass{sig-alternate-05-2015}

\usepackage{xcolor}

\usepackage{hyperref}
\usepackage{booktabs}
\renewcommand{\d}{\ensuremath{\mathrm{d}}}
\newcommand\myparagraph[1]{\textbf{#1}:}
\begin{document}

% Copyright
\setcopyright{acmcopyright}
%\setcopyright{acmlicensed}
%\setcopyright{rightsretained}
%\setcopyright{usgov}
%\setcopyright{usgovmixed}
%\setcopyright{cagov}
%\setcopyright{cagovmixed}

% DOI
%\doi{10.475/123_4}

% ISBN
%\isbn{123-4567-24-567/08/06}

%Conference
%\conferenceinfo{PLDI '13}{June 16--19, 2013, Seattle, WA, USA}

%\acmPrice{\$15.00}

%
% --- Author Metadata here ---
%\conferenceinfo{WOODSTOCK}{'97 El Paso, Texas USA}
%\CopyrightYear{2007} % Allows default copyright year (20XX) to be over-ridden - IF NEED BE.
%\crdata{0-12345-67-8/90/01}  % Allows default copyright data (0-89791-88-6/97/05) to be over-ridden - IF NEED BE.
% --- End of Author Metadata ---

\CopyrightYear{2016}
\setcopyright{acmcopyright}
\conferenceinfo{SE4Science'16,}{May 16 2016, Austin, TX, USA}
\isbn{978-1-4503-4167-7/16/05}\acmPrice{\$15.00}
\doi{http://dx.doi.org/10.1145/2897676.2897677}

\title{Nmag micromagnetic simulation tool -- software engineering lessons learned
  %\titlenote{(Produces the permission block, and
%copyright information). For use with
  %SIG-ALTERNATE.CLS. Supported by ACM.}
}
%%\subtitle{[Extended Abstract]
%%\titlenote{A full version of this paper is available as
%%\textit{Author's Guide to Preparing ACM SIG Proceedings Using
%%\LaTeX$2_\epsilon$\ and BibTeX} at
%%\texttt{www.acm.org/eaddress.htm}}}


\numberofauthors{3} %  in this sample file, there are a *total*
\author{
% 1st. author
\alignauthor
Hans Fangohr\\
       \affaddr{University of Southampton}\\
       \affaddr{SO17 1BJ Southampton}\\
       \affaddr{United Kingdom}\\
       \email{\small  fangohr@soton.ac.uk}
       % 2nd. author
\alignauthor
Maximilian Albert\\
       \affaddr{University of Southampton}\\
       \affaddr{SO17 1BJ Southampton}\\
       \affaddr{United Kingdom}\\
       \email{\small maximilian.albert@soton.ac.uk}\\
% 3rd author
\alignauthor
Matteo Franchin\\
       \affaddr{Cambridge}\\
       \affaddr{United Kingdom}\\
       \email{\small  franchinm@gmail.com}
}
       % There's nothing stopping you putting the seventh, eighth, etc.
% author on the opening page (as the 'third row') but we ask,
% for aesthetic reasons that you place these 'additional authors'
% in the \additional authors block, viz.
%\additionalauthors{Additional authors: John Smith (The Th{\o}rv{\"a}ld Group,
%email: {\texttt{jsmith@affiliation.org}}) and Julius P.~Kumquat
%(The Kumquat Consortium, email: {\texttt{jpkumquat@consortium.net}}).}
% Just remember to make sure that the TOTAL number of authors
% is the number that will appear on the first page PLUS the
% number that will appear in the \additionalauthors section.

\maketitle
\begin{abstract}
%We review design decisions and their impact for the open source code
%Nmag from a software engineering in computational science point of view. Key lessons to learn
%include that the approach of encapsulating the simulation
%functionality in a library of a general purpose language, here Python,
%eliminates the need for configuration files, provides greatest
%flexibility in using the simulation, allows mixing of multiple
%simulations, pre- and post-processing in the same (Python) file, and
%allows to benefit from the rich Python ecosystem of scientific
%packages. The choice of programming language (OCaml) for the
%computational core did not resonate with the users of the package (who
%are not computer scientists) and was suboptimal. The choice of Python
%for the top-level user interface was very well received by users from
%the science and engineering community. The from-source installation in
%which key requirements were compiled from a tarball was remarkably
%robust. In places, the code is a lot more ambitious than necessary:
%key routines work for $n$-dimensional space, but in simulations so far
%only 1d, 2d and 3d simulations were needed. Tests distributed with the
%package are useful, although more unit tests would have been
%desirable. The detailed documentation, together with a tutorial for
  %the usage of the system, was welcomed by the community.

  We review design and development decisions and their impact for the
  open source code Nmag from a software engineering in computational
  science point of view. We summarise lessons learned and
  recommendations for future computational science projects.  Key
  lessons include that encapsulating the simulation functionality in a
  library of a general purpose language, here Python, provides great
  flexibility in using the software. The choice of Python for the
  top-level user interface was very well received by users from the
  science and engineering community. The from-source installation in
  which required external libraries and dependencies are compiled from
  a tarball was remarkably robust. In places, the code is a lot more
  ambitious than necessary, which introduces unnecessary complexity and
  reduces maintainability. Tests distributed with the package are
  useful, although more unit tests and continuous integration would
  have been desirable. The detailed documentation, together with a
  tutorial for the usage of the system, was perceived as one of its
  main strengths by the community.
\end{abstract}





%
%  Use this command to print the description
%

% We no longer use \terms command


\section{Summary}
\label{sec-1}
\label{sec:summary}

The Nmag software was developed from 2005 to early 2012, and first
released in 2007 as open source code. Nmag has a Python user
interface, and uses Objective Caml code and established High
Performance Computing libraries under the hood, to solve
time-dependent partial differential equations using finite element
discretizations. It has been in use for nearly a decade, and with this review, we
try to share experience to help improve future generations of software
engineering projects in computational science.

\medskip
We start by summarizing key lessons and recommendations on this page below, with pointers to more detailed discussion in the main part of
the paper (Sect \ref{sec:introduction} to Sect \ref{sec:lessonslearned}).

\medskip

There are two points of view to consider in the development of a tool
for computational scientists. On the one hand, the \emph{end-users}
want the software to be easy to use in order to improve their
science. On the other hand, the \emph{developers} of the software care
about additional aspects, such as extensibility and
maintainability. We divide our recommendations according to these two
different points of view:

\noindent
\begin{center}
\begin{tabular}{p{0.84\columnwidth}c}\toprule
   \textbf{Recommendations primarily affecting end-users}& Sect. \\
\toprule
Embedding simulation into existing programming language provides unrivaled flexibility &\ref{sec:user-interface-through-python-library}\\\midrule
Python-based top-levels of the tool allow users to modify it to suit their needs
 &\ref{sec:choice-of-programming-language}\\\midrule
Python is a popular language that is perceived to be easy to learn by (non-computer) scientists &\ref{sec:choice-of-programming-language}\\\midrule

Documentation and tutorials are important &\ref{sec:documentation}\\
\bottomrule
\end{tabular}
\end{center}
\bigskip

\begin{center}
\noindent
\begin{tabular}{p{0.84\columnwidth}c}\toprule
\textbf{Recommendations primarily affecting developers}& Sect. \\\toprule
Version control tool use is essential &\ref{sec:version-control}\\\midrule
System tests are essential, unit tests are very useful &\ref{sec:testing}\\\midrule
Continuous integration is very useful &\ref{sec:installation}\\\midrule
Limit the supported or anticipated functionality to minimize complexity and enhance maintainability &\ref{sec:complexity}\\\midrule
Code generation based on user provided equations is up-front investment but widens applicability of tool &\ref{sec:code-generation}\\\bottomrule
Choice of unconventional programming language can limit the number of scientists joining the project as developers &\ref{sec:choice-of-programming-language}\\\midrule
OCaml not quite as fast as C/C++/Fortran &\ref{sec:ocaml-performance}\\\bottomrule
\end{tabular}
\end{center}

\section{Introduction}
\label{sec-2}
\label{sec:introduction}

Nmag is a micromagnetic simulation package. Micromagnetics is a
continuum theory of magnetization at the nano- and micrometer
scale. The magnetization is a continuous 3d vector field defined
throughout a ferromagnetic body. The dynamics of this vector field are determined
by an equation of motion, which depends on solving a non-linear partial
integro-differential equation. Using spatial discretization, this
partial differential equation (PDE) can be solved numerically. Nmag
uses a combination of finite element and boundary element
methods \cite[Section 2.2]{thesis-knittel} to
solve the integro-PDE in every time step. Within this finite element
model, the magnetization is represented as 3d-vector degrees of
freedom at every finite element node.  For the time
integration, the degrees of freedom at all nodes are treated as a
system of coupled ordinary differential equations, in line with the related
open source software Magpar \cite{scholz2003} and other tools that are
not freely available.

Nmag is summarized in a short paper \cite{fischbacher2007-nmag}.
More technical detail, in particular on the underlying multi-physics
finite element library \texttt{nsim}, is given in a manuscript available on
the arXiv \cite{fischbacher2009-nsim}. Some aspects of the parallel
execution model have been published
\cite{fischbacher2009-parallel}. The project home page
\cite{nmag-code-2012} contains the code as a tar-file. For reference,
the repository from which the tarball is built, is available as a
Mercurial repository on Bitbucket \cite{nmag-code-bitbucket}. A number
of PhD students have contributed to the software, and their theses
are an additional useful source of information, covering an overview
of the capabilities of the code \cite[chapter 5]{thesis-franchin},
long-range demagnetization field calculation, boundary elements and
matrix compression \cite{thesis-knittel}, macro-geometry periodic
boundary conditions and mesh generation \cite{thesis-bordignon}.

\bigskip

\begin{figure}
\centering
\includegraphics[width=0.65\columnwidth]{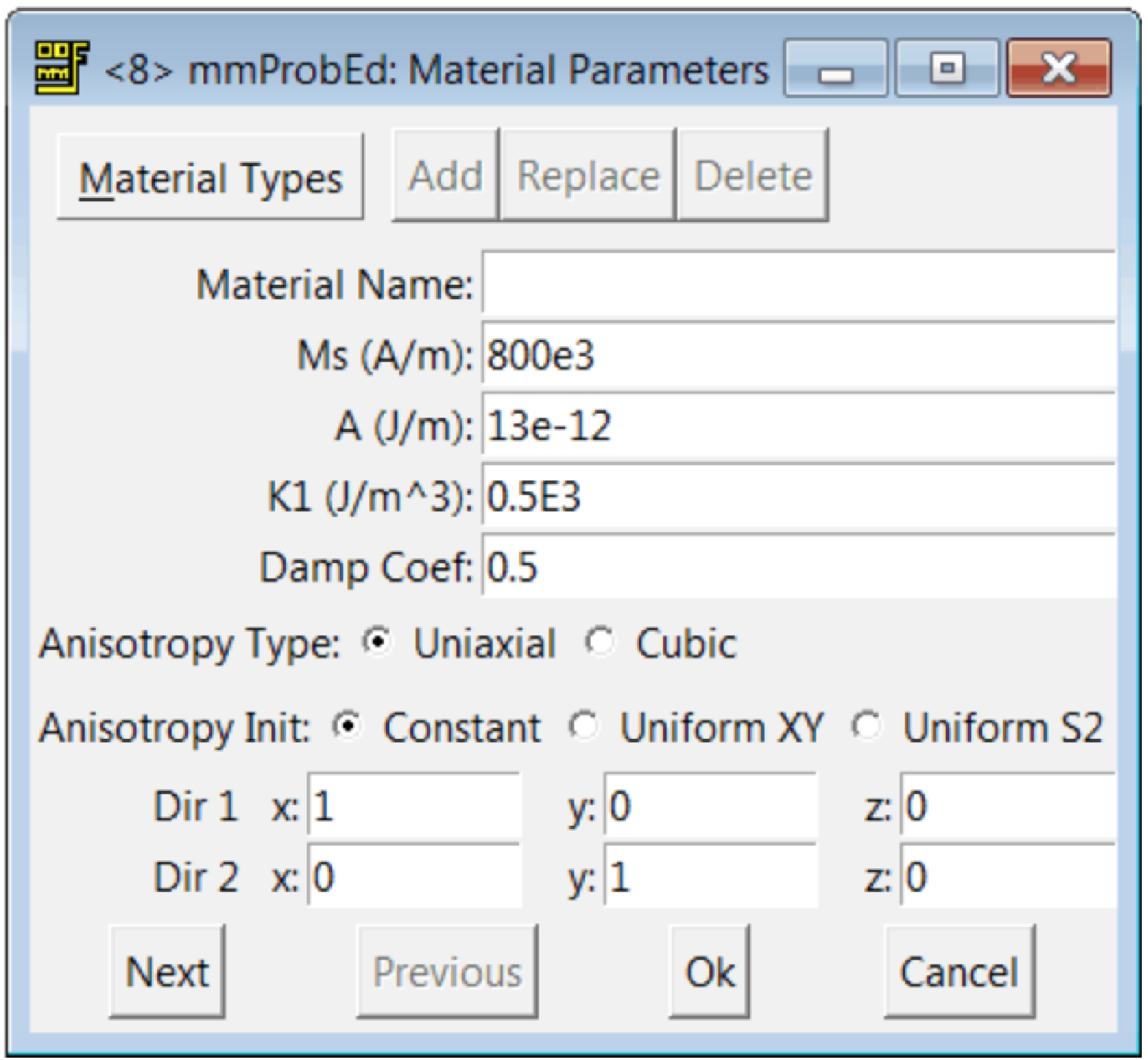}
\caption{The OOMMF graphical user interface for problem definitions \label{fig:oommf-problem-editor}}
\end{figure}

Computational micromagnetics became feasible and accessible with the
public domain release of the OOMMF Software by the National Institute
of Standards and Technology (NIST) in the late 90s
\cite{oommf1999}. The OOMMF software uses a finite difference
discretization (with different strengths and weaknesses than the finite
element discretization but outside the scope of this paper), which is
based on C++ code with a Tcl and Tk interface.
A simulation run can be configured through a graphical user interface (see
figure \ref{fig:oommf-problem-editor} for an example), or through
setting parameters in a configuration file which uses Tcl syntax.

The aim of the Nmag package was, among other things, to provide a
finite element based discretization approach to complement the OOMMF
tool. Simultaneously, and without being aware of each others
efforts, the finite element Magpar code was developed, and later
released as open source \cite{scholz2003}.

In this report, we focus on software engineering aspects of the
Nmag tool, including the user interface, the choice of languages and
tools, and the open source model. We summarize the Nmag project in
section \ref{sec:nmag}, before addressing the lessons learned in
section \ref{sec:lessonslearned}.

\section{Nmag project summary}
\label{sec-3}
\label{sec:nmag}

\begin{figure}
\centering
\includegraphics[width=0.9\columnwidth]{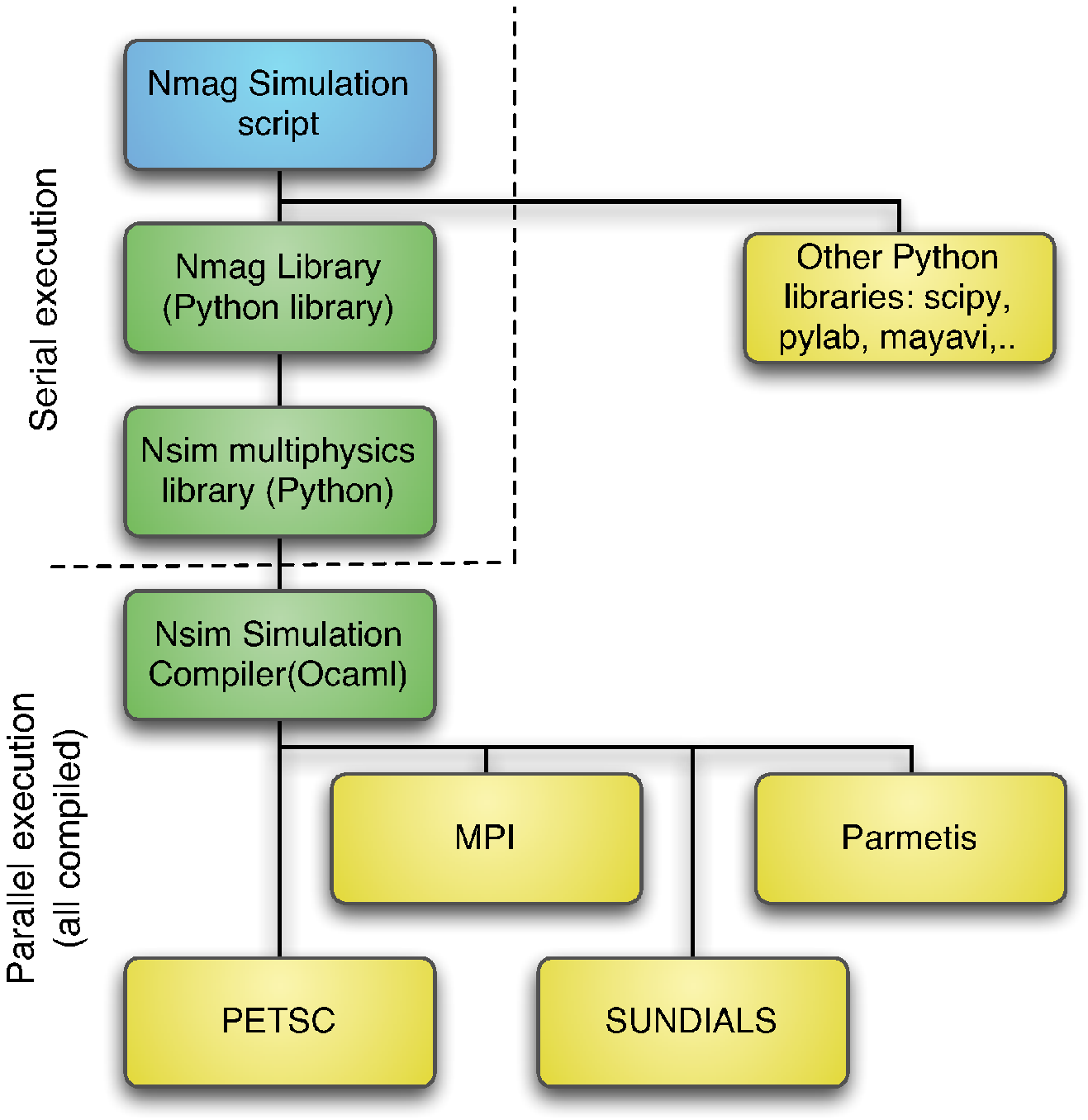}

\caption{Nmag architecture overview}
\label{fig:nmag-design}
\end{figure}

\myparagraph{Architecture} Figure \ref{fig:nmag-design} summarizes the Nmag architecture: the
blue box at the top represents the (Python) code that the end-user
assembles to define the simulation computation; this includes material
parameters, the file name of the finite element mesh to define the
geometry, the physical process that should be simulated etc. The
instructions are written as a Python script that makes use of the \texttt{nmag}
Python library (top green box), which in turn composes its
functionality from the (Python) \texttt{nsim} multi-physics library. The Python
Nsim multi-physics library is an interface to the functionality
provided through the implementation (bottom green box) in OCaml \cite{OCaml}. We
make use of existing high performance computing libraries such as
PETSC for sparse and dense matrix calculation,
PARMETIS for partitioning the mesh across multiple
MPI processes, and the CVODE time integration tools that come with the
SUNDIALS tool suite. The code is parallelized with MPI.

The code implements a dependency engine for physical fields
\cite[Section 4.4]{fischbacher2009-nsim}. This allows lazy
evaluation to only compute entities when they are truly
required, and to minimize the computation of fields that
depend on each others.

Periodic boundary conditions are difficult in micromagnetic
simulations due to the long range nature of the magnetostatic
interactions. A new computational model has been developed and
implemented (the 'macro geometry' \cite{fangohr2009-macrogeometry}),
which has subsequently been used by the Nmag successor code Finmag \cite{Finmag}, and other
micromagnetic tools such as the GPU-based package Mumax3
\cite{mumax3}.

\myparagraph{Time line} The planning for Nmag started around 2003, funding was secured in
2005, and the work started soon afterwards. The first version was
publicly released in 2007. The tool was actively maintained and
further developed until January 2012, when key developers moved on and
no further funding was available. Since then, the software has been
hosted `as is'.

\begin{figure}
\centering
\includegraphics[width=0.82\columnwidth]{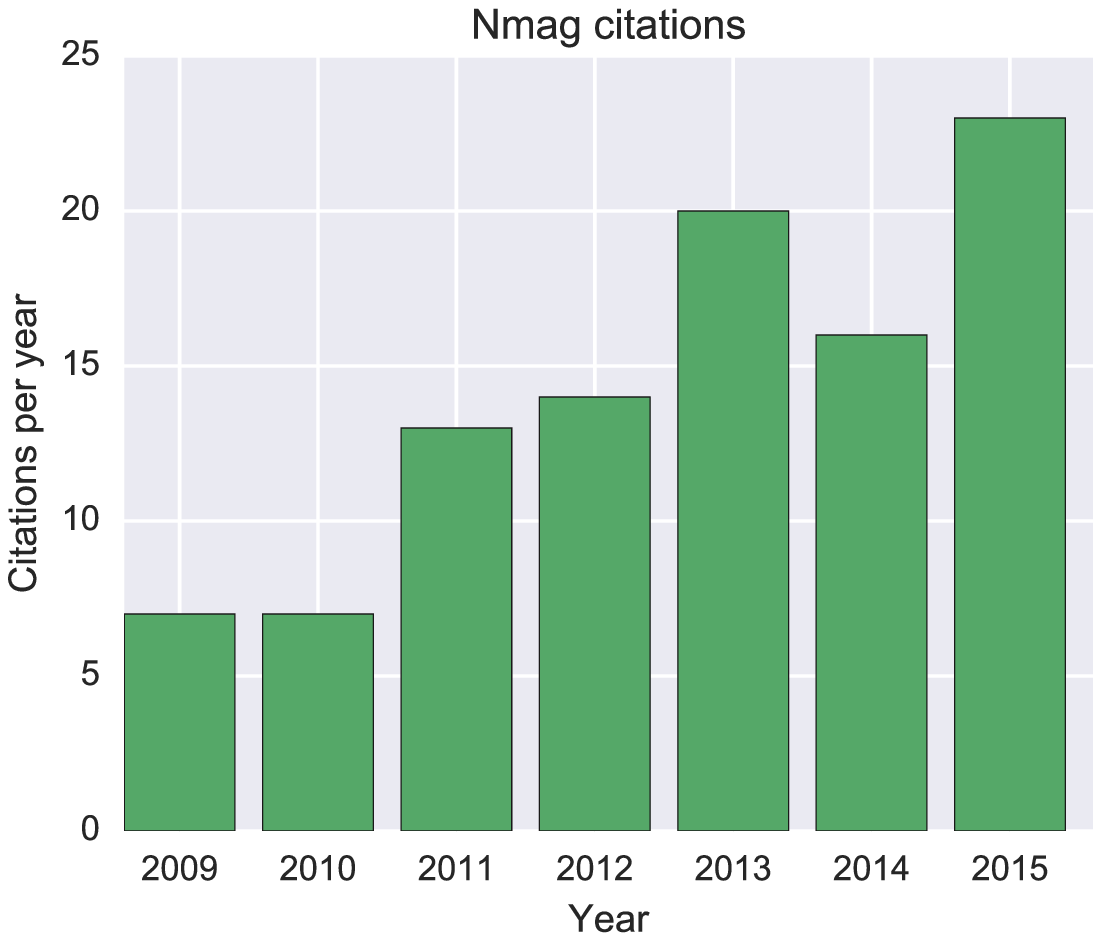}
\vspace{-0.3cm}
\caption{Nmag citations from Web of Science up to January 2016. }
\label{fig:nmag-citations}
\end{figure}

\myparagraph{Uptake} As of February 2016, the official Nmag publication
\cite{fischbacher2007-nmag} has been cited 103 times in publications
(Web of Science, Thomson Reuters) and 182 times on Google
Scholar. More than 150 users are known by name (from the mailing list,
or off-line queries), and the web site is frequented from academic and
industry domain names, with 45000 visits since 2006. Figure
\ref{fig:nmag-citations} shows the breakdown of citations per
year. Development and maintenance stopped early 2012, which (so far)
seems to not have affected the usefulness of the tool for research.

\myparagraph{Nmag team} The principal funding for the project was for a research fellow for
2.5 years. The team managed to find 3 PhD student projects that
supported the lead developer as part of their work
\cite{thesis-bordignon,thesis-franchin,thesis-knittel}, and in the
later phases there were smaller contributions from others. Later
(2009-2012), a research fellow from another project contributed
significantly to the project. The background of the team was in
physics and mathematics. None of the participants had any formal
training in software engineering, which is a common situation in
the development of research software.

\myparagraph{Community support} Community support involves the following tools and strategies: the
Nmag webpage hosts the code, installation instructions, the manual
\cite{nmag-code-2012}, a link to the mailing list archives
\cite{mailing-list-archives} and the Redmine Wiki. The University of
Southampton is hosting the mailing list, and Google Groups is used to
archive all communication in that mailing list. A Redmine server is
used simply to host a Wiki that users can edit
(\url{https://nmag.soton.ac.uk/community/wiki/nmag}).

\myparagraph{Software engineering process}
Parts of the software were written in an effectively plan-driven
approach, broken into separate requirement analysis, design,
implementation and testing phases. In particular the Nsim
multi-physics core was realized as one large piece of work by the lead
developer without significant subsequent change.

Other parts, in particular the Python-level micromagnetic interface
\texttt{nmag} were developed in a more agile style, with multiple
iterations of development, use of automated tests (section
\ref{sec:testing}), where both refactoring and additional feature
implementation was carried out in subsequent iterations.

\section{Lessons learned}
\label{sec-4}
\label{sec:lessonslearned}
\subsection{User interface through Python library}
\label{sec-4-1}
\label{sec:user-interface-through-python-library}
A key design decision was to embed the functionality of the simulation
into a general purpose language, in this case Python (see also
\cite[section 5.11.1]{fischbacher2009-nsim}).

\begin{figure}
\footnotesize
\includegraphics[width=0.9\columnwidth]{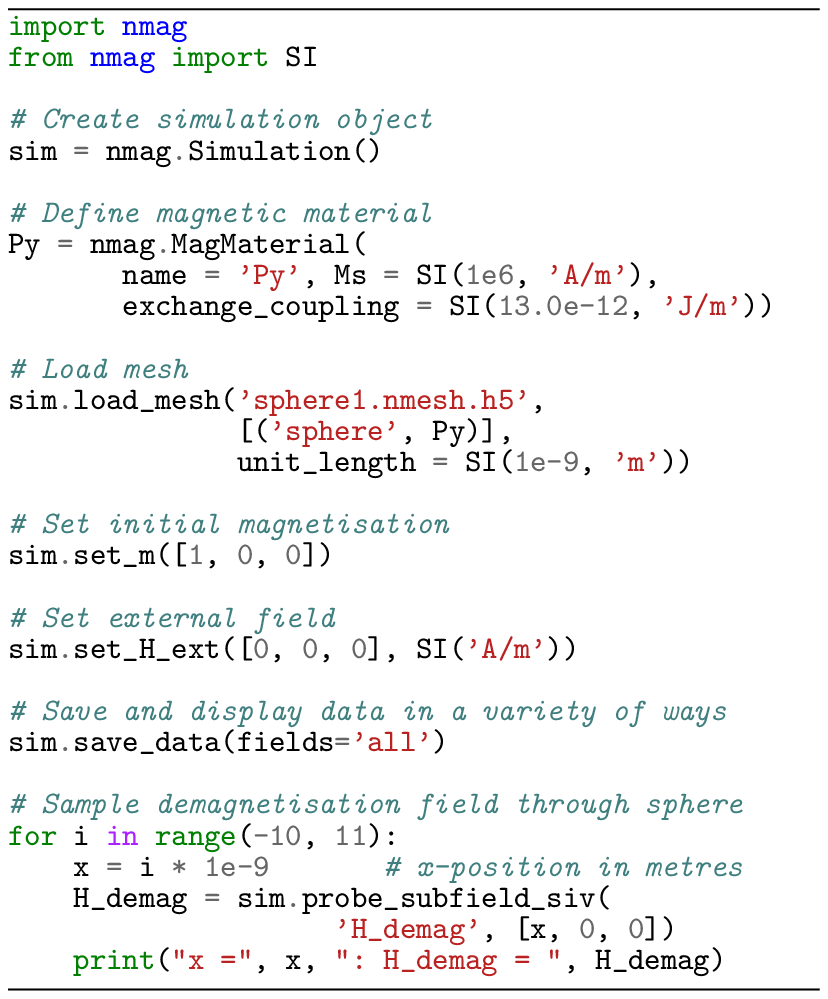}
\normalsize
\caption{Nmag end-user script example. \label{fig:nmag-code-example1}}
\end{figure}

Figure \ref{fig:nmag-code-example1} shows an Nmag simulation script,
which is a Python script that imports and uses the \texttt{nmag}
library.

%\inputminted[bgcolor=white,frame=lines]{python}{figure4.py}
%\input{figure4.pygtex}
%\lstinputlisting{figure4.py}

In comparison to definition of simulation configurations through
configuration text files or graphical user interfaces (see for example Figure~\ref{fig:oommf-problem-editor}), this approach has a number of
advantages: (i)~no parser needs to be written -- Python is the parser,
(ii)~the user has complete freedom in using Python constructs
to combine the simulation commands provided by the \texttt{nmag} library
as needed for the particular application,
(iii)~data pre- and post-processing, and calculations that take place
during the simulation can make use of the Python ecosystem of
available scientific libraries, (iv)~the configuration of the
simulation can make use of Python functions to provide, for example,
initial magnetization vector field configurations, that show a
complicated spatial dependence. (See also section 5.11.2 in
\cite{fischbacher2009-nsim}).
\subsection{Choice of programming languages}
\label{sec-4-2}
\label{sec:choice-of-programming-language}

Figure \ref{fig:cloc-output} shows some statistics regarding the
programming languages / tools and respective lines of code in the Nmag
project.

Nmag uses Python as the language for the user to define how the
micromagnetic simulation should be carried out. It should be noted
that the micromagnetic capabilities as well as the testing
infrastructure and system tests are implemented at the Python level,
while the OCaml code provides a generic multi-physics simulation
environment and contains significant parts of the multi-physics finite
element code. The OCaml multi-physics engine is capable of solving
problems in areas outside micromagnetism.
%, for example electric
%resistivity \cite{amr-resistance} and heat conduction
%\cite{jouleheating}.

Python was a good choice for the high level user-interface, and
internals of the package: it is a user-friendly language
\cite{fangohr2004comparison} that has gained substantially in
popularity in computational science and elsewhere since 2005;
resulting in additional benefits from the critical mass of
users. Anecdotal evidence and the increase of Python-based simulation tools and libraries support our point that Python is a high productivity language in computational
science. In micromagnetism, packages developed after Nmag have
followed the model of providing a Python library to offer their
functionality \cite{MagnumFE,Finmag}.

For Objective Caml (OCaml) as the work horse of the multi-physics
engine, the situation is less clear. There are technical reasons why
OCaml is a good choice, including its C interoperability interface,
its (interactive) interpreter and its native code compiler. OCaml also
offers automatic memory management, expressive power and functional
capabilities (see section 5.13 in \cite{fischbacher2009-nsim}).

In hindsight, we have identified a disadvantage of a social
engineering nature: OCaml is not a wide-spread language (certainly not
outside computer science and mathematics), and there are virtually no
users of the Nmag software that have OCaml experience. Furthermore,
OCaml is very rarely taught and due to its (powerful but somewhat
non-mainstream) functional style, it presents a steep learning curve
to the typical Nmag user, who tends to be a researcher in material
science, engineering, physics, biology, geography and medicine (but
generally not a computer scientist). For an open source project, it is
important that the code is accessible by the user community to attract new contributors.

Comparison of execution performance of compiled OCaml code with C/C++ code
have shown that the OCaml code can be noticeably slower than the C
code. We address this point in more detail in
section \ref{sec:ocaml-performance}.

\begin{figure}
\begin{center}
\begin{tabular}{lrrrr}
Language & files & comment lines & code lines\\
\hline
OCaml & 174 & 15111 & 53445\\
Python & 588 & 17718 & 49286\\
C & 49 & 2548 & 12375\\
Bourne Shell & 47 & 1232 & 9184\\
make & 138 & 391 & 2831\\
C/C++ Header & 14 & 410 & 820\\
\hline
SUM: & 1010 & 37410 & 127941\\
\end{tabular}
\end{center}
\vspace{-0.3cm}
%\footnotesize
%\inputminted[bgcolor=white]{text}{figures/cloc-output.txt}
%\normalsize
\caption{Output of \texttt{cloc} (v1.65) run on Nmag source code, tests, and documentation files.\label{fig:cloc-output}}
\end{figure}
\subsection{Python interpreter activated from OCaml}
\label{sec-4-3}
For technical reasons and the capabilities of the existing PyCaml
interface \cite{pycaml} that connects OCaml and Python, the Nmag setup
is such that the Python interpreter is called from within an OCaml
executable.

In more detail: the users starts an executable called \texttt{nsim}
compiled from OCaml code. This executable initializes the simulation
engine, and then calls an embedded Python interpreter, which processes
the user's simulation commands (which are typically given through a
file \texttt{mysim.py} to the \texttt{nsim} executable, \emph{i.e.} the
command line call would be \texttt{nsim mysim.py}). If no file to
process is given, a Python interpreter is
displayed in which Python commands can be entered interactively, and
the \texttt{nmag} library is accessible.

An alternative setup would be that the user starts the 'usual' Python
interpreter, and that a \texttt{nmag} package can be imported, which
carries out the housekeeping work, initialization of MPI and and
execution of required OCaml code when imported.

With hindsight, we suggest that the alternative arrangement would have
been much preferable for a number of reasons: (i) \texttt{nsim}
activates a Python interpreter (in which the \texttt{nmag} library is
accessible), so it will seem more logical to the user if that command
is \texttt{python} and not \texttt{nsim}. (ii) The Python interpreter
coming through \texttt{nsim} is in general different from the system
Python interpreter (and with the current install, see section
\ref{sec:installation}, this Python interpreter is built from source
and installed locally); and third party Python modules (such as
\texttt{numpy}, \texttt{scipy}, \texttt{matplotlib}) may end up having
to be installed separately for both interpreters.

\subsection{OCaml performance}
\label{sec-4-4}
\label{sec:ocaml-performance}

The OCaml-based multi-physics engine subdivides the computation into two
phases: initialisation and time integration. In the first phase, a set of
sparse and dense matrices are pre-computed and stored in memory. These
matrices capture the various physical interactions that govern the
behavior of the simulated system and are mainly computed by OCaml
code. In the second phase, the time integration is carried out, using
the pre-computed matrices. These computations are mainly done using the
PETSC library.
Micromagnetic simulations of real materials and devices require fine
meshes corresponding to pre-computed matrices of size of tens of
Gigabytes and matrix assemble times of the order of hours. It is
therefore important to make the OCaml initialisation phase fast and
efficient. Investigations of the OCaml engine code performance revealed
some limitations of the language and its compiler when used in our
context, which eventually led us to rewrite parts of the
matrix initialization code in C.

First we comment on array efficiency. OCaml native arrays are
unidimensional. A rank-2 tensor (i.e. a matrix) can still be represented
as an array of arrays of floats which, however, is not memory efficient:
the floats are not stored in contiguous areas of memory, requiring
extra memory for storing pointers and thus additional indirections
during array accesses. Moreover, the sub-arrays are not guaranteed to
have the same lengths. This makes multi-dimensional arrays difficult to
analyze and optimize for a compiler. While the \texttt{Bigarray} module
in OCaml provides multi-dimensional arrays which are stored contiguously
in memory and thus allow to overcome some of these problems, accesses to
OCaml big-arrays are not inlined, leading to poor
performance \cite{ocaml-performance}.

Second, we have found that OCaml has limited support for some
compiler optimisations that are particularly useful in numerical code, such as
bounds-checking elimination, loop unrolling, and vectorisation. These
optimization techniques are now common in mainstream languages and are
performed by freely available compilers such as \texttt{gcc} and
\texttt{clang}, which of course receive a vastly greater investment
and contribution from the private software industry.

We provide electronic supplementary material \cite{ocaml-performance}
with sample code that underpins the results reported in this section,
and additional interpretation. The examples show that rewriting an
OCaml loop in C/C++ can give speedups of a factor 4. This is
significant because performance critical numerical code often consist
of simple for-loops.

In summary, the lack of native multi-dimensional array support is
a problem for scientific code, and a better compiler would bring
OCaml closer to C in terms of performance.
\subsection{Symbolic derivation of PDE calculations at run-time}
\label{sec-4-5}
\label{sec:PDEs}

In finite elements, a mathematical problem in form of a PDE is solved
in a given number of spatial dimensions $n$; these are typically
$n=3$, or $n=1$, $n=2$, corresponding to 3d space as we know it, and
reduced models where a 1d or 2d space is sufficient. On these
$n$-dimensional domains, we operate with scalars, vectors or tensors,
which can have their dimensionality $k$. Finally, there is a variety
of basis functions and (taking only continuous Galerkin elements as an
example), these have their own polynomial order $p$.

Traditionally, the right equation for a particular mathematical
operator is derived with pen and paper for particular values of the
dimensions of space $n$, the dimensionality of the degree of freedom
$k$ and basis function order $p$ . Once the equation has been derived,
it is hard-coded as an implementation for this specific case. This
code is then used to populate the finite element matrices. In the
micromagnetic context, the Magpar package \cite{scholz2003} is such an
example.

Nmag's approach is different: here, the relevant analytic operations,
which include differentiation and integration, are carried out
symbolically (within the Nsim OCaml code base) to generate specialized
instructions to compute the matrix entries for the particular
operator, dimensionality, basis function order etc. that the user
requires. The \texttt{nsim} code supports arbitrary basis function orders $p$,
dimensionality of the degree of freedom $k$ and and arbitrary
dimensionality $n$ of the domain.

This approach provides much greater flexibility to change the equations
in the model (an important consideration in the context of exploratory research),
or the numerical model parameters such as
the order of the basis functions. It also avoids repetitive manual
analytical work, and -- assuming the symbolic computations are
implemented correctly -- reduces the chance of errors. We discuss the
associated additional complexity in section
\ref{sec:arbitrary-number-of-dimensions}.

\subsection{Auto-generation of code for local field mappings}
\label{sec-4-6}

\label{sec:code-generation}

The primary entities of interest in finite element simulations are \emph{fields},
such as the magnetization vector field, the temperature scalar field,
a displacement vector field, etc.
Nmag uses auto-generation of code at run time to allow the user to
compute tailored expressions that map from one field to another field
(details in \cite[section 5.11.4]{fischbacher2009-nsim}).
There are two possible ways of achieving this:

(i) The user provides C code that contains the mathematical mapping
operation that is required.  At run time (as the user-provided string
is not know before then), this user-provided C-code string is embedded
into a template that provides access to the relevant field arrays and
access methods, and the combined C-code is written to disk, compiled, and
dynamically linked. The auto-generated code will automatically translate
the user provided indices to the right memory locations, which is
non-trivial for multi-physics simulations where multiple fields are
defined at every node.

(ii) The user can also provide algebraic expressions which represent the
required operation, which are automatically translated into C code. We
provide an example to demonstrate this. In the micromagnetic problem,
there is a magnetization vector field $\mathbf{m}(\mathbf{r})$ that
defines a 3d vector at every point $\mathbf{r}$ in 3d space. In this
example, we look at the mapping of the magnetization $\mathbf{m}$ onto
its time derivative $\frac{\d \mathbf{m}}{\d t}$ as is necessary to
compute the equation of motion (\ref{eq:eom}) for this magnetization
vector function $\mathbf{m}(\mathbf{r})$:
\begin{equation}
\frac{\d \mathbf{m}}{\d t} = c_1 \mathbf{m}\times\mathbf{H} + c_2\mathbf{m} \times (\mathbf{m}\times\mathbf{H}) \label{eq:eom}
\end{equation}
We note that $\mathbf{H}$ is obtained as a function of $\mathbf{m}$
by solving certain partial differential equations \cite[Section 2.2]{thesis-knittel}.

We can rewrite (\ref{eq:eom}) using index notation as:
\begin{equation}
\frac{\d m_i}{\d t} =
\sum_{j, k} \left[
  c_1 \epsilon_{ijk}m_jH_k +
  \sum_{p, q} c_2 \epsilon_{ijk}m_j (\epsilon_{kpq}m_p H_q) \label{eq:index}
\right]
\end{equation}

For this formulation, \texttt{nsim} provides a small domain specific language. We show how
equation (\ref{eq:index}) is represented as a string in this domain specific language:

{\footnotesize
\begin{verbatim}
dmdt = """%range i:3, j:3, k:3, p:3, q:3
  dmdt(i) <-   c1 * eps(i, j, k) * m(j) * H(k)
             + c2 * eps(i, j, k) * m(j)
             * eps(k, p, q) * m(p) * H(q)"""
\end{verbatim}
}

We have found the interface (ii) useful to quickly and flexibly extend
the equation of motion. The ability to specify C code directly through
method (i) allows to sidestep the \texttt{nsim} framework where functionality
is required that was not anticipated initially.  By using C code
(rather than Python code, say) good performance is achieved, in
particular when loops over all degrees of freedom are involved.
\smallskip
A very similar mechanism to method (i) is provided in the multi-physics
finite element library FEniCS \cite{fenics} through
the \texttt{instant} module, which can be used to initialize fields with
arbitrary C expressions. An approach similar to (ii) is used in FEniCS
for the non-local PDE operators.

FEniCS \cite{fenics} started being developed at the same time as Nsim and is
now widely used, including
in the micromagnetic package Magnum.fe \cite{MagnumFE} and the Nmag
successor software Finmag \cite{Finmag}. FEniCS has core routines
written in C++ and provides a Python interface.

\subsection{Parallel execution model}
\label{sec-4-7}
The parallel execution model of Nmag is that only one Python process
is running, driving slaves through MPI from within the OCaml code
\cite{fischbacher2009-parallel}. This allows end-users to write truly
sequential Python code and to completely ignore the parallel execution
of the micromagnetic equations. This is different from the FEniCS
parallel model \cite{fenics}, where also the Python code executes in
parallel. While the FEniCS model requires more thought at the Python
level, it allows to add computationally demanding operations to be
executed in parallel through Python. For expert users and scalability,
the FEniCS model is preferable.

\subsection{Complexity originating from generality}
\label{sec-4-8}
\label{sec:complexity}
\subsubsection{Arbitrary number of dimensions}
\label{sec-4-8-1}
\label{sec:arbitrary-number-of-dimensions}

As introduced in section \ref{sec:PDEs}, in finite elements, some
mathematical problem is solved in a given number of spatial dimensions
$n$; where the \texttt{nsim} code supports calculations in arbitrary number of spatial dimensions.

While there are problems defined on space that is higher dimensional
than 3d space in science and engineering, none of that functionality
has been used in the lifetime of the Nmag software. The complexity of
the code could have been reduced (and maintainability and
accessibility increased) if we had limited its functionality to
spatial dimensions $n$ of 1, 2 and 3.

We note for context that the FEniCS \cite{fenics} multi physics
library follows a similar path of using symbolic calculation at
run time to derive finite element matrix entries, and that the FEniCS
functionality is limited to 3 or fewer dimensions.

\subsubsection{Arbitrary high level language support}
\label{sec-4-8-2}

The operator notation used in (the Python) \texttt{nsim} library is
representing differential operators through a string (see section 4.3.1 and
example A.2 in \cite{fischbacher2009-nsim}). An alternative would be
to create classes in Python that represent mathematical objects and
differential operators, and use operator overloading to integrate the
mathematical description naturally within the Python language. The
FEniCS project \cite{fenics} has followed the latter path with their
Python interface \cite{dolfin}.

The motivation behind the design decision to prefer the string
representation over the (object-oriented) operator representation for
the differential equation was to avoid coupling the notation too
tightly to the Python scripting language: by sticking to strings, one
can substitute Python by another language more easily as and when required.
Note that when Nmag was created Python was by far
not as mainstream in scientific computing as it is today; for example,
the Python numerical library \texttt{numpy} was only created in 2005.

From a user's point of view, we believe that it is preferable to use
Python objects over strings to define the differential equations: this
is more natural and allows to exploit auto-completion and to explore
capabilities and documentation of objects when working
interactively with the Python prompt.

It may have been beneficial to approach this design question in a more agile
way by fully buying into the Python language and the overloaded
operator notation initially and revisiting the decision as the
project evolved. As a decade has passed, we now know that
there is no need to introduce another high level language for Nmag:
Python is doing fine, and the development of the Nmag project has
stopped anyway. Another possibility would be to introduce a bridge
element that translates the object oriented equations into strings
when required, which would allow combining the long term strategy
of expressing differential equations through strings with the ability
to offer operator overloading in Python to the user.

\subsubsection{Multi-physics capabilities}
\label{sec-4-8-3}

Nmag is built on the multi-physics library Nsim, and thus the
micromagnetic model that is available through Nmag is only one of
many possible types of PDE-based simulations that could be built
on top of \texttt{nsim}. The multi-physics capabilities were indeed unique in the micromagnetic
simulation landscape, but bring significant additional complexity.

While we attempted to immediately implement a multi-physics framework,
and then build a micromagnetic simulation on top of this, it may have
been more efficient to build a micromagnetic prototype first, and use
the experience gathered with this in developing a more generic
framework.

We note that as the multi-physics capabilities were not part of the
funded research program, they have not been developed and documented
to the level of the core micromagnetic functionality, and are thus not
used widely.

\subsubsection{Associating numbers with units}
\label{sec-4-8-4}
All Nmag input physical entities have to be expressed as a product of
a number and product of powers of dimensions (such as m, kg, s, A, K,
mol, cd). For example, the Python expression
\texttt{x = SI(100e-9, "m")}
describes a length of 100 nano meters. For multi-physics
simulations, this provides great advantages as the output fields
emerging from user-defined operations automatically carry the right
units. It also allows to scale numbers internally (to reduce chances
of overflow etc), and to use arbitrary unit systems (such as SI, CGS,
or custom). In the context of micromagnetics -- Nmag's application domain -- the capability met a
mixed reception: some users find it useful, others dislike the
overhead of having to type the dimensions, when they normally use SI
units (as is enforced in the OOMMF \cite{oommf1999} package).
\subsection{Version control and source}
\label{sec-4-9}

\label{sec:version-control}

Development of Nmag started in 2005 and has used three different
version control systems through its history: starting with CVS, before
changing to Subversion, before switching to Mercurial
in 2010. Initially, the source code was included in the tarballs
containing the 'source' releases of the project, and later the
Mercurial repositories have been made available on Bitbucket
\cite{nmag-code-bitbucket}. There are separate repositories for the
source, the tests, the documentation, the webpage, the distribution,
and one additional meta repository that provides a script to clone
them all together into the right relative subdirectory structure. Use
of version control tools is essential. The distributed tools, such as
Mercurial and Git, are more flexible than CVS and Subversion.
\subsection{Testing}
\label{sec-4-10}
\label{sec:testing}
There about 75 tests coming with the code, combining unit tests (of
only some parts of the code), with a fair number of system tests, and
a few regression tests. While undoubtedly more unit tests would have
been desirable, the availability of the existing tests is extremely
useful as a first indicator of a working installation,
etc. Test-driven development was not generally used for the project.
\subsection{Documentation}
\label{sec-4-11}

\label{sec:documentation}

In addition to documentation in the source code files, there is the
official Nmag documentation that is available \cite{nmag-code-2012} as
\href{http://nmag.soton.ac.uk/nmag/current/manual/singlehtml/manual.html}{html}
or \href{http://nmag.soton.ac.uk/nmag/current/manual/manual.pdf}{pdf}
(183 pages). Key components include a tutorial-like, step-by-step
introduction and walk-through of Nmag, starting from simple and common
use cases to more complicated and specialized application examples
(112 pages), explanation of general concepts (7 pages), a command
reference, which is built from the Python documentation strings
defined in the source code (20 pages), an overview of executable
scripts including their options and usage examples, and Nmag data file types
(12 pages), a list of 20 frequently asked questions and answers (9
pages), a mini-tutorial into micromagnetic modeling (not specific for
Nmag, but experience shows that new Nmag users are also often new to
the field). The documentation is often cited by users as being very good, and useful.

\subsection{Installation}
\label{sec-4-12}

\label{sec:installation}

Nmag depends on a large number of support libraries. Many of these are
of scientific origin and change often, including changes in the
interface. As a result, the Nmag code needs updating to compile
correctly after any such update.

For some time, there were installation options through a Debian
package, a KNOPPIX Live Linux CD (this was before virtual machines were widely
used), and from source.

Later, we focused our energy on providing an installation setup that
is as independent as possible from version changes of libraries that
are installed through the Linux distribution to maximize the chances
for long term availability in a situation without maintenance
resources.

The resulting setup compiles and installs all the required
dependencies\footnote{\scriptsize atlas-3.6.0, cryptokit-1.2, gsl-1.14, ipython-0.1, lapack, mpich2-1.2.1p1, numarray-1.5.2, numpy-1.5.0, ocaml-3.12.0, ocamlgsl-0.6.0, ocaml-findlib-1.2.1, ParMetis-3.1.1, Python-2.7.2.tar.bz2, petsc-lite-3.1-p5, PyVTK, qhull-2003.1, tables-2.1.2, scipy-0.7.2 py-0.9.1 ply-3.3, sundials-2.3.0, HLib-1.3p19}
from source, then builds Nmag based on these support libraries.
Compilation does only work on Linux (although on a large
variety of distributions and releases).

This setup has been remarkable robust, and only failed once in 4
years of no maintenance updates. The one failure was due to more
recent \texttt{gcc} versions reporting commenting through a double slash in C
header files of the hdf5 library as illegal, which previously only triggered a warning.

The price to pay for this robustness is that the compressed tarball
containing all the library source code is about $91\,\mathrm{MB}$ in size. Once
Nmag has compiled all the support libraries, these take together $1.4\,\mathrm{GB}$
of disk space storage. After removing temporary build files $0.5\,\mathrm{GB}$ remain.
As the support libraries are snapshots
from their source distribution in early 2012, no improvements or
bug fixes developed in the support libraries will affect the Nmag
compilation and executables.

It would have been useful to use continuous integration (to run tests,
build documentation and releases) to support more frequent releases.

\medskip
\emph{Acknowledegments}: We acknowledge financial support from the UK's Engineering and
Physical Sciences Research Council (EPSRC) from grants EP/E040063/1,
EP/E039944/1 and Doctoral Training Centre EP/G03690X/1), from the
European Community's FP7 Grant Agreement no. 233552 (DYNAMAG), from
Horizon 2020 Research Infrastructures project \#676541 (OpenDreamKit),
and from the University of Southampton, all of which contributed in
part to the development of the Nmag tool suite and this review.
\bibliographystyle{abbrv}
\bibliography{paper.bib}  % paper.bib is the name of the Bibliography in this case
% ACM needs 'a single self-contained file'!
%\balancecolumns
\end{document}